# Room-temperature Continuous-wave Lasing from Monolayer Molybdenum Ditelluride with a Silicon Nanobeam Cavity

Yongzhuo Li[1], Jianxing, Zhang[1], Dandan Huang[1], Hao Sun[1], Fan Fan[1,2], Jiabin Feng[1], Zhen Wang[1], and C. Z. Ning[1,2,*]

[1]Department of Electronic Engineering, Tsinghua University, Beijing, 100084, China

[2]School of Electrical, Computer, and Energy Engineering, Arizona State University, Tempe, Arizona 85287, USA

*Correspondence to: cning@asu.edu

**Abstract**

Monolayer transition metal dichalcogenides (TMDs) provide the most efficient optical gain materials and have potential for making nanolasers with the smallest gain media with lowest energy consumption. But lasing demonstrations based on TMDs have so far been limited to low temperatures. Here, we demonstrate the first room-temperature laser operation in the infrared wavelengths from a monolayer of molybdenum ditelluride on a silicon photonic-crystal nanobeam cavity. Our demonstration is made possible by a unique choice of TMD material with emission wavelength below silicon absorption, combined with the high $Q$-cavity design by silicon nanobeam. Lasing at 1132 nm is demonstrated at room-temperature pumped by a continuous-wave laser, with a threshold density at 6.6 W/cm$^2$. The room-temperature linewidth of 0.202 nm is the narrowest with the corresponding $Q$ of 5603, the largest observed for a TMD laser. This demonstration establishes TMDs as practical nanolaser materials. The silicon structures provide additional benefits for silicon-compatible nanophotonic applications in the important infrared wavelengths.

**One Sentence Summary**

We report the first room temperature operation of a silicon nanobeam cavity laser integrated with monolayer molybdenum ditelluride in the near infrared wavelength.

Monolayer transition metal dichalcogenides (TMDs) have attracted a great deal of attention recently due to their distinctive electronic and optical properties and potential applications in integrated optoelectronic devices(*1–4*). One of the most important properties is the transformation from indirect to direct band-gap semiconductors when the thickness of TMDs is reduced down to a monolayer(*5, 6*), making monolayer TMDs perfect choice as the thinnest optical gain materials. Another important property is the large exciton binding energy of TMDs well beyond thermal energy at the room temperature(*7–10*), providing a unique opportunity for room temperature lasing on much stronger excitonic transitions (than the electron-hole plasma transitions) or even exciton-polariton lasing. Therefore, monolayer TMDs provide an efficient optical gain material, especially attractive for making nanolasers with the smallest volume of gain medium. While nanolasers have been an active field of research in its own right(*11–21*), the lack of an efficient gain medium with the smallest possible volume limits the ultimate performance for energy efficient applications and the overall device sizes(*22*). Such nanolasers are attractive for a wide range of applications such as coherent light sources for future on-chip integrated photonic systems, or for flexible displays, in addition to a wide range of interests in fundamental exciton-polariton physics. Several TMD-based nanolaser demonstrations have been reported at cryogenic temperatures using a monolayer $WSe_2$ coupled onto a gallium phosphide (GaP) photonic crystal cavity([1]), or $WS_2$ with a silicon nitride ($Si_3N_4$) microdisk([2]). The main reason for low temperature operation seems to be the relatively low cavity $Q$ factor (measured at ~1300([1]) at room temperature, 2465([1]) and 2604([2]) at cryogenic temperature). Emission into cavity modes has also been reported at room temperature from multi-layer $MoS_2$([23], [24]) with the $Q$ factor in the range of 1000 to 3300. Lack of room temperature lasing demonstration not only hinders possible practical applications, but also raises a more serious question whether such ultimately thin gain material is fundamentally capable of providing enough optical gain for practical optical applications which invariably require operation at room temperature. In addition, all the emission-related device demonstrations so far have been limited to the visible wavelengths. For the most important potential applications such as on-chip optical interconnects, infrared wavelengths are necessary, especially below Si-absorption band. Lasing demonstration in the infrared is in general more challenging due to typically weaker emission and less sensitive detection.

There are a few interrelated challenges for realizing a monolayer-TMD based laser at the room temperature. Foremost is the design and fabrication of a high $Q$ cavity, which so far has been limited to around 3000 due to the limit of material choices. Silicon provides the best material bases as a cavity material due to its high index of refraction and the maturity in the fabrication techniques. Silicon based 1D photonic crystal nanobeam cavity has produced $Q$ as high as $7.5 \times 10^5$ ([25]). But TMDs materials that have received most of the attention for optical study (such as S- and Se-based TMDs $WS_2/Se_2$ or $MoS_2/Se_2$) have emission wavelengths that are strongly absorptive for silicon. Monolayer molybdenum ditelluride ($MoTe_2$) has a bandgap around ~1.72 eV([9]), but an excitonic photoluminescence (PL) emission peak of ~1.1 eV([26]) is about 50 meV below silicon bandgap and thus nearly transparent for silicon([27]). Silicon's weak absorption at 1.1 eV is estimated at 1.5 1/cm([28], [29]) and silicon is thus perfectly suited for silicon nanobeam cavity for $MoTe_2$. Importantly the choice of silicon has an additional advantage of being compatible with a host of silicon photonics functionalities and thus a room temperature silicon nanobeam TMD nanolaser in near IR wavelengths could be potentially important for future on-chip integrated silicon photonics.

In this article, we report the first room temperature operation of a silicon nanobeam cavity laser integrated with a monolayer of $MoTe_2$ in the near IR wavelength below the silicon absorption under the pumping of a continuous-wave He-Ne laser. The details of design and fabrication are described in the Supplementary Materials([27]). The material and the cavity structure are shown in Fig. 1. The monolayer $MoTe_2$, with a thickness estimated to be 0.7 nm([27]) was exfoliated from a commercial bulk $MoTe_2$ (supplier: 2D Semiconductors) as shown in Fig. 1A and 1B. The schematic, optical, and scanning electron microscope (SEM) images of the device structure are shown in Fig. 1C, 1D, and 1E, respectively.

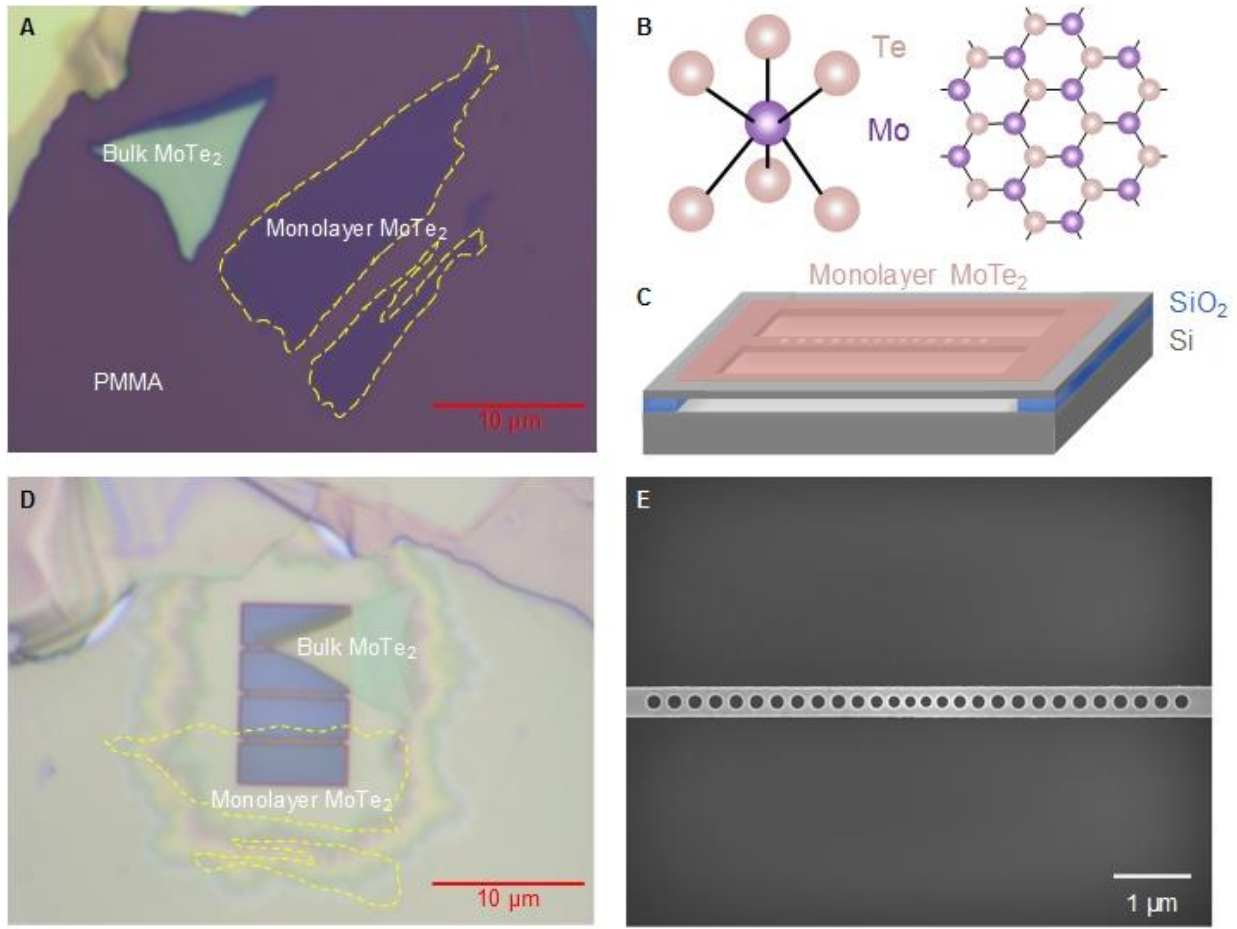

**Fig. 1. Monolayer MoTe$_2$ and silicon photonic crystal nanobeam cavity**. (**A**) Optical image of bulk (greenish region) and monolayer MoTe$_2$ (contoured region) on PMMA. (**B**) Crystal structure of MoTe$_2$ (2H). (**C**) Schematic of our device (the Si photonic crystal nanobeam laser structure suspended in air with a monolayer MoTe$_2$ on top). (**D**) Optical image showing fabricated nanobeam laser with the transferred monolayer MoTe$_2$ as marked by the yellow dashed lines (There are three silicon nanobeam cavities in the image, but only the bottom one was covered with monolayer MoTe$_2$). (**E**) SEM of an undercut silicon nanobeam cavity. The dimensions of nanobeam cavity is 7.2 μm (length)×0.36 μm (width) ×0.22 μm (thickness) (see also Fig. 2A for details).

The nanobeam cavity of 365 nm in width is fabricated using the top layer (with a thickness of 220 nm) of a silicon-on-insulator (SOI) chip(*27*), as shown in Fig. 1E. Our high $Q$ nanobeam cavity, as schematically shown in Fig. 2A, consists of mainly three segments(*25*): two "mirror" sections on either side and a "cavity" section in the middle. The mirror section has 10 periods of holes of 94 nm in diameter with a pitch of 268 nm. The middle cavity section contains 8 air holes with diameter and period gradually decreasing by a step of $0.057r_n$ and $0.057a_n$ from the side towards the middle (see Fig.2A). The two side sections of periodic arrays of air holes serve as two mirrors of a cavity due to the photonic bandgap (PBG) effect(*30–33*), while the symmetrically size-tapered middle section is designed to reduce radiative loss of the cavity(*25*, *34*). The structural parameters of the nanobeam cavity are crucial and optimized for achieving high $Q$ values of cavity modes. Vertical to the beam axis, optical field is confined by the high refractive index silicon beam suspended in air. Figures 2B-2D show the electric field ($E_y$) in the X-Y plane of the first three modes, with transverse electric (TE) polarization. The calculated wavelengths of these three modes are 1054, 1132 and 1167 nm, respectively, which agree very well with the measured mode wavelengths in the PL spectrum, as shown in Fig. 2E. The calculated $Q$-values for the three modes are, $5.2 \times 10^6$, $6.5 \times 10^5$, and $1.4 \times 10^3$, respectively. The second mode at 1132 nm is expected to be the lasing mode for three reasons: (1) The wavelength of second mode is close to peak wavelength of PL spectrum; (2) The $Q$ factor of the second mode is, while slightly smaller than the first mode, still very high; (3) The second mode (1.5 cm$^{-1}$) has smaller Si-absorption than the first mode (16.3 cm$^{-1}$)(*29*). Thus our designs focus on the second mode as the lasing mode and the wide mode separation avoids mode competition from the first and third modes.

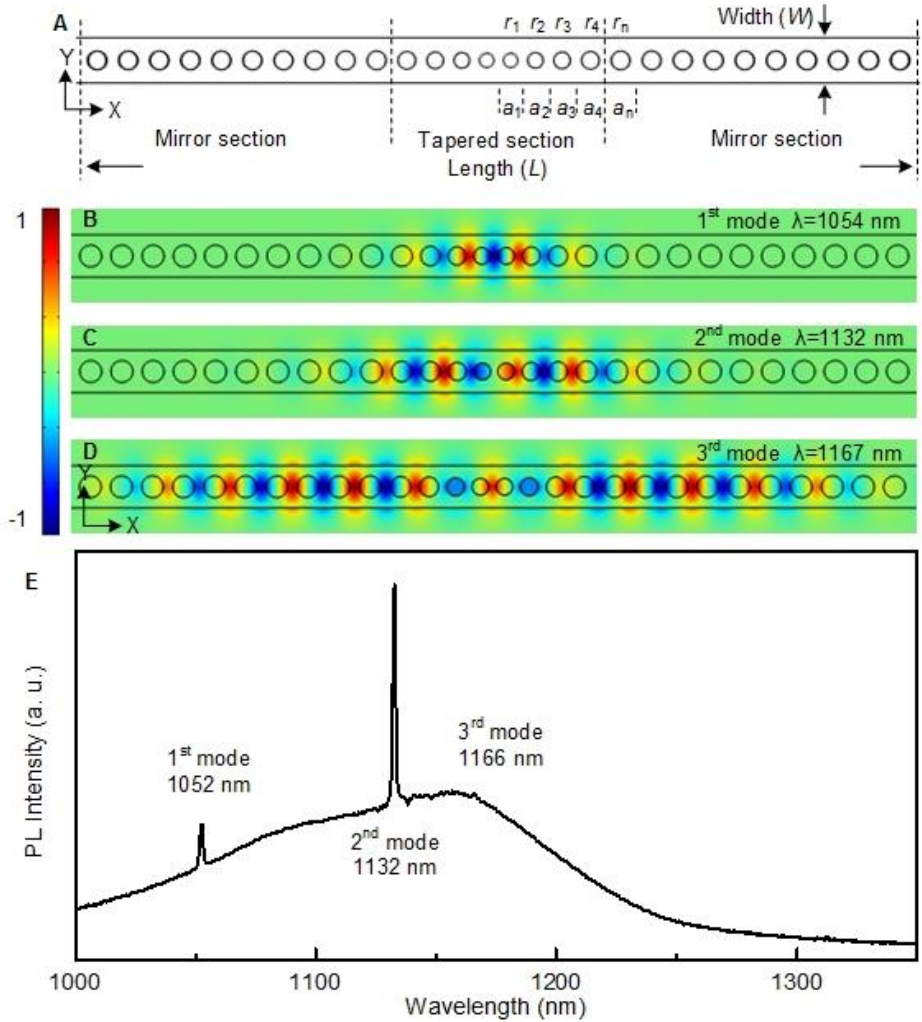

**Fig. 2. Design and optical modes of the photonic crystal nanobeam cavity**. (**A**), Schematic of the nanobeam cavity. (**B**)-(**D**) Electric field $E_y$ component in X-Y plane of the first three modes. (**E**) PL spectrum of the nanolaser at room temperature under the pump power above the lasing threshold.

All measurements were performed at room temperature pumped by a 633-nm CW laser (He-Ne lasers, Thorlabs)(*27*). Figure 3A shows the emission spectra under different pump levels, showing clearly the emergence of a strong laser peak at 1132 nm with increasing pump. At the low pump level of 54.8 µW, only broad-band spontaneous emission spectrum is observed. As the pump is further increased, we see the occurrence and gradual increase of lasing modes of the second modes. The continued increase of broad spontaneous emission background can be explained as follows: The area of active monolayer $MoTe_2$ and the size of the pumping beam are both much larger than that of nanobeam cavity, especially in the direction normal to the beam axis. Since the spontaneous emission from the region of monolayer $MoTe_2$ far away from the nanobeam cavity is not coupled to the cavity modes, they mainly contribute to the background PL spectrum, leading to a continuous increase of PL away from the lasing peaks. This is different from a typical semiconductor laser where one expects a depressed spontaneous emission background above the threshold due to carrier density clamping. The light-in vs light-out curve in a log-log plot is shown in Fig. 3B, which clearly shows the transition from spontaneous emission to lasing in the form of a typical S-curve. The slope of over 3 in the middle section indicates the rapidly amplifying spontaneous emission and the occurrence of a clear laser threshold.

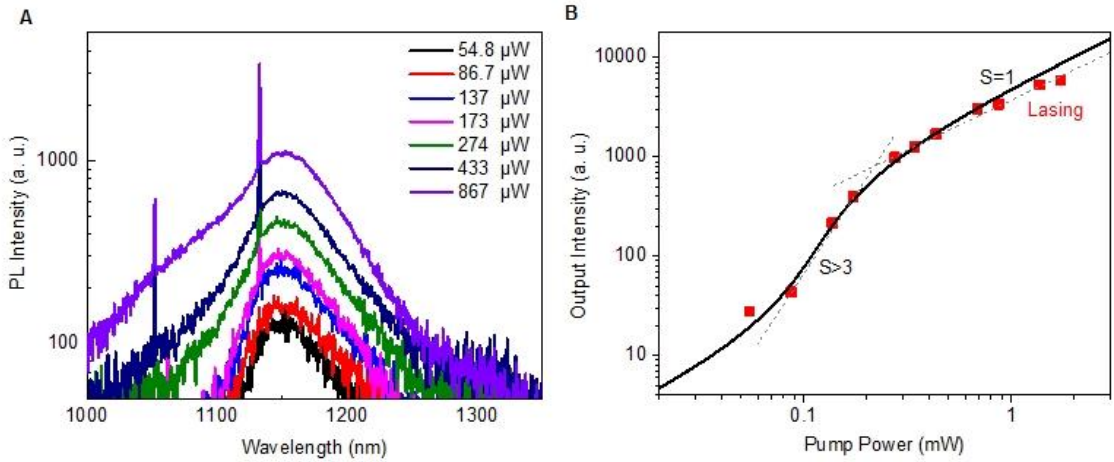

**Fig. 3. Room temperature emission**. (**A**) PL spectra of nanobeam laser with increasing pump power levels at room temperature, using a grating of 150 g/mm, corresponding to an estimated spectral resolution of 0.41 nm. (**B**) Log-log plot of light-in vs light-out, showing a clear transition from spontaneous emission via amplified spontaneous emission to eventual lasing with the associated scaling power of larger than 3 in the transition region and eventually back to 1 above the threshold, where the light-out intensity is obtained by removing the broad-band spontaneous emission background. Solid squares represent experimental measurement, while solid line is a rate-equation fitting. The light-dotted line indicates the slopes of experimental data points.

To identify the laser characteristics more accurately, PL measurement was conducted at a high resolution estimated to be 0.09 nm at increasing pump levels, as shown in Fig. 4A with a clear lasing peak at 1132 nm, consistent with the result presented in Fig. 3. The full-width at half-maximum (FWHM) of PL spectrum at 173 µW pump is only 0.202 nm corresponding to a *Q* value of 5605 for the operating wavelength of 1132.25 nm. This is the highest *Q* value of any 2D TMD-based laser demonstrated so far. Figure 4C is the light-in vs light-out curve in a log-log plot, extracted from the data in Fig. 4A, with the solid squares represent the intensity around 1132 nm. There is a clear nonlinear region with the scaling index estimated to be larger than 2, signifying a laser threshold.

The experimental data is fitted by solving laser rate equations(*27*) with the spontaneous emission factor as a fitting parameter. As is seen, $\beta$ =0.1 is the best fitting to the experimental data as compared with other curves for $\beta$=0.05, 0.2, and 0.9. The high *Q* factor of the nanobeam cavity significantly enhances the spontaneous emission through the Purcell effect. The Purcell factor was calculated from the simulation and the value for the second mode is 76(*27*). The large spontaneous emission factor effectively couples emission into the high *Q* nanobeam cavity. These are crucial factors for achieve lasing operation at room temperature in our case. It is worth mentioning that the smooth transition as a result of the large $\beta$ makes the laser threshold somewhat less well defined and the maximum of the slope gives a threshold(*35*) of 0.097 mW, corresponding to a power density of 6.6 W/cm$^2$(*27*). This value is much lower than those for any other semiconductor lasers at room temperature, indicating potential important role of nanolasers based on 2D materials as energy efficient lasers for future on-chip interconnects.

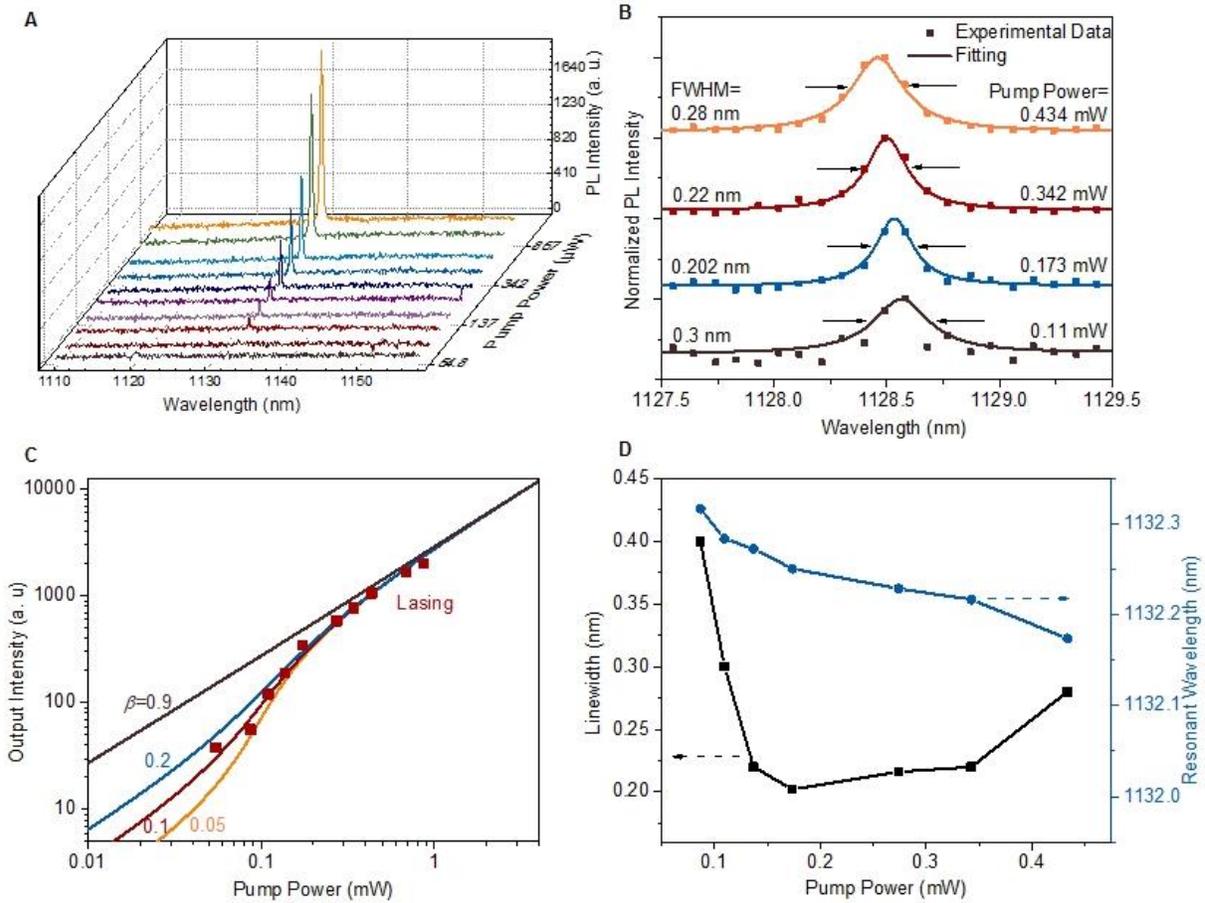

**Fig. 4. Lasing characteristics at room temperature**. (**A**) PL spectra at increasing pump levels using a grating of 600 g/mm, corresponding to a spectral resolution of 0.09 nm. (**B**) The fitting of PL spectra under different pump levels. The solid squares are the measured data. The lines represent the Lorentz fittings. (**C**) Log-log plot of light-in vs light-out where solid squares are the experimental data. Solid lines are results of a rate-equation calculation(*27*) with different spontaneous emission factor $\beta$. $\beta$=0.1 is the best fitting to the experimental data. (**D**) Linewidths and resonant wavelengths versus pump power.

The experimental emission peaks are fitted to Lorentzian line-shape and shown in Fig. 4B. The FWHM of the cavity mode is extracted through Lorentzian fitting and shown in Fig. 4D. The linewidth shows a significant reduction from 0.4 nm to 0.202 nm when pump power increases from 0.087 to 0.173 mW. Assuming that 0.4 nm corresponds to the "cold cavity" linewidth, we obtain a cold-cavity $Q$ of 2830, much smaller than the theoretical $Q$ from our design, due to fabrication imperfection and the presence of an overlayer of PMMA on top of MoTe$_2$ layer. The range of pump power for linewidth narrowing matches well with "kinked" region or laser threshold of Fig. 4C. We notice a significant linewidth re-broadening with the further increase of pumping above the threshold, which is often seen in other lasers too for reasons ranging from heating effect to carrier density increase. In our devices, absorption of the pump laser by the Si-nanobeam cavity leads to a high carrier density in Si-nanobeam. Such carrier-density increase leads to decrease in the refractive index and increase of absorption coefficient of Si at the modal wavelength of nanobeam cavity(*36*). The change (the decrease) of refractive index leads to blue-shift of the cavity modes of the nanobeam cavity(*27*). The estimated absorption increase for resonant wavelength is around 8 per cm under the pump power of 0.434 mW. Such increase of absorption is equivalent to a cavity $Q$ decrease, leading directly to the linewidth broadening. The linewidth is broadened to 0.251 nm through calculation, very close to the measured linewidth of 0.28 nm under the pump power of 0.434 mW. We emphasize that it is essential to fabricate high quality nanobeam cavity with a high $Q$ to achieve lasing at room temperature. We also transfer monolayer MoTe$_2$ onto nanobeam cavities with rough sidewall. Only cavity-enhanced spontaneous emission peaks were observed in the PL spectra.

Some of the quantitative numbers are worth mentioning. The total modal volume for the lasing mode is calculated to be only 0.48 $(\lambda/n)^3$ (*27*), which represents one of the smallest nanolasers. The volume of the active gain medium is estimated to be 2440 nm×0.7 nm×365 nm= $(85.4 \text{ nm})^3$. Such a modal volume is likely the smallest volume of the active gain medium of any room temperature lasers ever realized and could have significant impact on low threshold and energy efficient lasers for on-chip applications. The overall device size is 7200 nm×365 nm×220 nm= $(833 \text{ nm})^3$, or $(0.73\lambda)^3$, important for integrated photonics where small footprint is a necessity. This small volume of active region also leads to an extremely low threshold. The threshold power density of 6.6 W/cm$^2$ at room temperature is much lower than that of room temperature excitonic laser in UV wavelengths(*37*).

In summary, we have demonstrated the room-temperature operation of a monolayer semiconductor nanolaser for the first time. The impacts of this demonstration can be understood in several ways: First, room temperature operation of semiconductor lasers of any kind has historically represented important milestones, since such capability is a necessary first step toward any practical applications. Second, monolayers of TMDs allows the fabrication of photonic devices with the smallest possible volume of gain medium and device sizes, but the lack of a room temperature demonstration has generated a general doubt about the feasibility of such smallest optical gain medium overcoming the total loss. This demonstration will thus contribute the establishment of 2D TMDs as a practically viable gain medium and will open a range of novel lasers and photonic devices. Third, since our nanobeam lasers operate on excitonic transitions well below the bandedge of MoTe$_2$, our demonstration indicates the possibility of an excitonic lasing at room temperature, which is still a challenging task in infrared regime and deserves to be further studied. The room temperature realization(*37*) in UV wavelength shows a threshold that is higher than in our case. Furthermore, a collection of favorable quantitative measures such as small overall device sizes, small modal volume, small active volume, and extremely low threshold power density could lead to exciting applications in high energy efficiency on-chip interconnects. The Si-nanobeam cavity would make such an application more appealing for photonics integration. Finally, our combined

MoTe$_2$-nanobeam cavity structure can potentially open a wide range of investigations at room temperature such as electrical injection 2D TMDs-based lasers, valley-spin polarized lasers, or strong cavity-TMD monolayer coupling physics.

## Acknowledgements

This research is supported by the 985 University Project of China and Tsinghua University Initiative Scientific Research Program (No. 20141081296). The authors thank Prof. Yidong Huang for the usage of their fabrication equipment.

C.Z.N. initiated the research on the silicon based monolayer MoTe$_2$ lasers, and supervised the overall project. Y.L. developed the simulations and design of devices. Y.L., J.Z., D.H. and J.F. exfoliated monolayer MoTe$_2$ from bulk material. Y.L., J.Z. and D.H. fabricated the devices. Y.L., J.Z., H.S. and F.F. performed the optical measurements and data analysis. J.Z. and Z.W. carried out the laser equation fitting. Y.L. and C.Z.N. analyzed data and wrote the manuscript. All authors participated in the discussions.